\documentclass[aps,prl,reprint,superscriptaddress,longbibliography]{revtex4-1}

\usepackage[utf8]{inputenc}
\usepackage[english]{babel}
\usepackage{amsmath}
\usepackage{amssymb}
\usepackage{graphicx}
\usepackage{bm}
\usepackage[colorlinks = true,
            linkcolor = blue,
            urlcolor  = blue,
            citecolor = blue,
            anchorcolor = blue]{hyperref}

\begin{document}
\title{Collective dynamics of domain walls: an antiferromagnetic spin texture in an optical cavity}

\author{Jephthah~O. \surname{Iyaro}}
\affiliation{Department of Physics and Astronomy, University of Manitoba, Winnipeg, Manitoba R3T 2N2 Canada}
\email{iyaroo@myumanitoba.ca}

\author{Igor \surname{Proskurin}}
\affiliation{Department of Physics and Astronomy, University of Manitoba, Winnipeg, Manitoba R3T 2N2 Canada}
\email{Igor.Proskurin@umanitoba.ca}
\affiliation{Institute of Natural Sciences and Mathematics, Ural Federal University, Ekaterinburg 620002, Russia}

\author{Robert~L. \surname{Stamps}}
\affiliation{Department of Physics and Astronomy, University of Manitoba, Winnipeg, Manitoba R3T 2N2 Canada}
\email{Robert.Stamps@umanitoba.ca}

\date{\today}

\begin{abstract}
Spin canting and complex spin textures in antiferromagnetic materials can often be described in terms of Dzyaloshinskii-Moriya interactions (DMI). Values for DMI parameters are not easily measurable directly, and often inferred from other quantities. In this work, we examine how domain wall dynamics in an antiferromagnetic optomagnonic system can display unique features directly related to the existence of DMI. Our results indicate that the presence of DMI enables spin interactions with cavity photons in a geometry which otherwise allows no magneto-optical coupling, and on the other hand modulates frequencies in a geometry where coupling is already realized in the absence of DMI. This result may be used to measure the DMI constant in optomagnonic experiments by comparing resonances obtained with different polarisations of the exciting field.
\end{abstract}

\maketitle

\section{Introduction}
Cavity optomagnonics in which precision measurements of interactions between photons and magnons are studied \cite{viola2016} has potential applications in quantum information processing \cite{Hisatomi2016} and spintronics \cite{Karenowska2014}. Essentially, it allows a description where the force on magnetic moments appears as an optical pressure of photons. We discuss here an optomagnonic analog of cavity optomechanics \cite{Aspelmeyer2014} in which one is able to observe coupling of electromagnetic radiation to spin texture oscillations. \cite{Brahms2010,viola2016}. As a consequence, one is able to realize optomechanical phenomena in magnonic systems, which includes mode attraction \cite{bernier, harder}.

Cavity magnonics has been explored primarily for ferromagnetic systems, with frequencies in the GHz range \cite{goryachev, tabuchi, zhang2014,huebl,Zhang2016a,haigh,Osada2016}. 
At present, antiferromagnets are receiving attention as possible platforms for exotic spin textures such as skyrmions, which can exist in low-dimensional materials with broken inversion symmetry.
Antiferromagnets are characterized by zero net magnetization, dynamics in the THz range \cite{baltz} and typically have large magnetic anisotropies that make them useful for spintronic applications, but at the same time makes difficult experimental observation of collective magnetic dynamics.   
Here, we propose to use magneto-optical coupling to enable collective motion of antiferromagnetic textures, which is a well established experimental technique to study ultrafast dynamics in antiferromagnetic insulators \cite{tzschaschel}.

The Dzyaloshinskii-Moriya interaction (DMI) can exist in non-centrosymmetric magnetic material and is known to play an important role in describing non-trivial spin textures. In one spatial dimension DMI leads to such things as long-periodic modulated structures known as soliton lattices, while in two spatial dimensions DMI helps to stabilize topologically non-trivial skyrmion lattices. It is also known to play an important role in domain walls \cite{janutka,muratov}, where it affects the DW structure and lifts degeneracy between DWs with opposite chiralities. In this paper, we show that DMI can also affect how spin textures are coupled to optical modes.

Whereas some of the most interesting textures are found in two and three dimensions, there have been some studies on one-dimensional structures such as solitons in antiferromagnets \cite{braun2005,ichiraku,kolezhuk}. 
Here, we examine the essential features for a domain wall. Domain walls are one-dimensional topological textures whose dynamics can be described in analogy to that of massive particles. The collective motion of ferromagnetic and antiferromagnetic domains walls is well understood \cite{bouzidi,tretiakov,tveten,tveten2,tveten3,kim_2014,bang}. It has also been studied for other spin textures such as soliton lattices in helimagnets \cite{bostrem,kishine}.

Coupling of collective magnon modes to optical cavity modes have been realized in a number of experiments \cite{haigh,tabuchi,quirion,Osada2016}. Recently, a model for cavity magnonics with a ferromagnetic domain wall was proposed and parameters for the coupling were estimated \cite{proskurin_2019}. In this case the frequency for the ferromagnetic domain was in the GHz range for domain wall oscillations in a pinning site.

In this paper, we discuss a cavity optomagnonic system where an antiferromagnetic domain wall is coupled to optical photons via the inverse Faraday effect. This leads to the non-linear optomechanical-type coupling where the domain wall feels a pressure from cavity photons.  We highlight features such as the hybridization of optical and magnonic modes, and identify a level attraction regime, which is characterized by the coalescence of the real parts of the eigenfrequencies of two modes in the region marked by an exceptional point \cite{heiss2004}. This behaviour has been attributed to instability in optomechanical systems due to negative frequency in the effective Hamiltonian of an externally driven system \cite{proskurin2,bernier}.

We investigate the effects of DMI on domain wall dynamics in two different geometries. We find that in one geometry where DMI favours one chirality of the domain wall, the presence of DMI enables coupling of magnons to photons which is otherwise nonexistent  without DMI. In a second geometry, the DMI exerts a torque which leads to domain wall tilt and consequent effects on the coupling terms. We expect that these features are not limited to the case of a single domain wall and can be found in other chiral spin textures. This can assist detection of DMI in thin films of antiferromagnetic materials.


\section{The model}
\begin{figure}
  \centerline{\includegraphics[width=8.6cm]{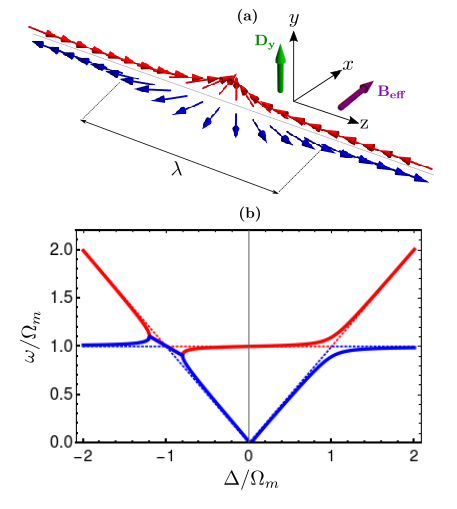}} 
\caption{(a) Schematic illustrating the geometry where the effective magnetic field of the light along $\hat{x}$ interacts with the domain wall with width $\lambda$ and axis along $\hat{z}$ in the presence of DMI along $\hat{y}$. (b) Hybridized frequency of the cavity modes and magnon modes of the antiferromagnetic domain wall as a function of the detuning parameter in the absence of DMI. In the absence of DMI, there is no coupling achieved as indicated by the dotted lines. The thick lines represent what happens in the presence of DMI when the magnon modes couple to the cavity modes. The real part of the two modes attract in the region of negative detuning and repulsion is observed in the region of positive detuning.}
\label{fig:fig1}
\end{figure}

Our model begins with a N\'{e}el domain wall in a one-dimensional antiferromagnet as shown in Fig. (\ref{fig:fig1}~a). The red and blue arrows denote the magnetization direction on two magnetic sublattices. The sublattices A and B have opposite spins, $S_{\text{A}}$ and $S_{\text{B}}$, respectively. The preferred magnetization rotation plane is the $x$-$z$ plane. The Hamiltonian describing this system is:
\begin{multline}\label{1}
\mathcal{H}_{\text{dw}} = \sum_{\langle i,j \rangle}J  \vec{S}_i \cdot \vec{S}_j  - K_z \sum_i (\vec{S}_{i,z})^2 - \sum_i K_x (\vec{S}_{i,x})^2 \\
+ \sum_i (-1)^{i+1} \vec{D} \cdot (\vec{S}_i \times \vec{S}_{i+1}).
\end{multline}  
The first term on the right is the nearest-neighbour exchange interaction between spins. The second and third terms are the easy and intermediate axis anisotropies in the $z-$ and $x-$ directions, respectively \cite{conzelmann} with constants $K_z \gg K_x$. The last term is the antisymmetric interfacial DMI whose alternating sign prevents a spiral spin state. \cite{Papanicolaou2}

We introduce the total and staggered magnetization, $\vec{m}_i = (\vec{S}_A^i + \vec{S}_B^i)/2S$ and $\vec{l}_i= (\vec{S}_A^i - \vec{S}_B^i)/2S$, respectively subject to the constraints that $\vec{m} \cdot \vec{l}= 0$ and $\vec{m}^2 + \vec{l}^2 =1$ \cite{papanicolaou,tveten,tveten2,lan}. It is useful make a continuum approximation, $\vec{m}_{i+1} \approx \vec{m}_i + a_0 (\partial \vec{m}_i /\partial x)+ ... $, where $a_0$ is the lattice constant. We parameterize the staggered magnetization, $\vec{l}$ by polar and azimuthal angles $\theta$ and $\phi$ in spherical coordinates: $\vec{l}=(\sin \theta \cos \phi, \sin \theta \sin \phi, \cos \theta)$. By these definitions, the Hamiltonian of the system written in the continuum approximation is:
\begin{multline}\label{2}
\mathcal{H}_{\text{dw}}= \int_{-\infty}^{\infty} \frac{dz}{a_0}\left[ A \left(\frac{\partial \theta}{\partial z} \right)^2  - K_x \sin^2 \theta \cos^2 \phi 
\right.
\\
\left.
- K_z \cos^2 \theta
- D_x \left(\frac{\partial \theta}{\partial z} \right) \sin \phi + D_y \left(\frac{\partial \theta}{\partial z} \right) \cos \phi \right],
\end{multline}
where $A=J/2 $ is the exchange constant, $D_x$ and $D_y$ are the $x$ and $y$ components of the DMI, respectively (See Appendix \ref{APP:DMI} for the derivation of the DMI).

\begin{figure*}
\centerline{\includegraphics[width=1.0\textwidth]{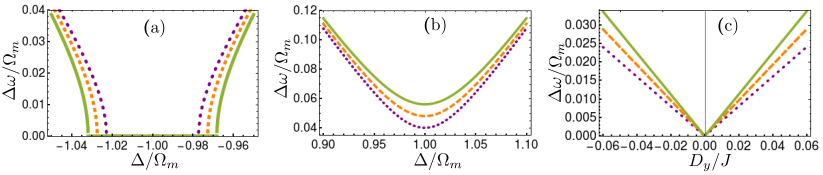}}
\caption{(a) Frequency gap in the presence of DMI of strength $0.06 J$ as a function of the negative detuning around the point of attraction for different coupling strengths: $g=0.10$, $g=0.12$, and $g=0.14$ represented by the purple dotted, orange dashed and green solid lines, respectively. Around the exceptional point, attraction dominates a wider range for larger coupling strengths and decreases for lower coupling strengths.  (b) Frequency gap in the presence of DMI of strength $0.06J$ as a function of the positive detuning. For the same value of detuning, larger frequency gaps are observed for larger coupling. In all cases, the gap is minimum at resonance and increases as we move away from resonance. (c) Frequency gap as a function of negative and postive DMI present along the $\hat{y}$. At $D_y=0$ there is no gap. In the presence of DMI, the frequency gaps increase linearly with increasing magnitude of DMI and the gap is observed to be larger for larger coupling strengths.}
\label{fig:fig2}
\end{figure*}


The domain wall static profile is obtained using the standard methods \cite{conzelmann,coey_2010,braun,tveten,lan}. The domain wall tilt angle is defined as $\phi$ with respect to the $x$-$z$ plane and assumed to be constant. The energy density is minimized via a Euler-Lagrange equation. The wall profile is $\theta(z) = 2 \tan^{-1} \exp (z-z_0/\lambda)$, where $\lambda = \sqrt{A/(K_z-K_x \cos^2 \phi)}$ is the characteristic domain wall width. Our analysis of the static wall profile dependence on DMI agrees with results in Ref. \cite{conzelmann}. We summarize the key details in subsequent paragraphs.

The DMI-dependent wall energy, $\mathcal{E} = 4 \sqrt{A(K_z - K_x \cos^2 \phi)} - D_x \pi \sin \phi + D_y \pi \cos \phi$  is obtained by substituting the wall profile into Eq.~(\ref{2}) and integrating. Since the energy density depends on only the $x$ and $y$ components of the DMI, two possible geometries are considered.

First, DMI is considered to be present in the $\hat{x}$ direction. This results in a distortion of the spin orientation on the sublattices and consequently results in a DMI-dependent tilt angle,
 \begin{equation}\label{33}
 \phi(D_x) = \text{sgn} D_x \cdot \cos^{-1} \left(\pm \sqrt{\frac{(8 J K_x^2 - D_x^2 K_z \pi^2)}{(8 J K_x^2 - D_x^2 K_x \pi^2)}}\right),
 \end{equation}
which minimizes the energy density. This dependence remains valid until a saturated DMI strength of $D_{x,\text{sat}}= 4/\pi \sqrt{J K_x^2 / 2 K_z}$ is reached beyond which $\phi(D_x) = \pi/2$. For the second geometry where DMI is present along $\hat{y}$, the DMI rather breaks the degeneracy of the energy minima and favours one chirality of the wall at a critical value, $D_{y,\text{c}}= 4 K_x/\pi \sqrt{J/2(K_z-K_x)}$. The tilt angle, $\phi(D_y)=0$ if $D_y<D_{y,\text{c}}$. Otherwise, $\phi(D_y)=\pi$.


\begin{figure}
  \centerline{\includegraphics[width=8.6cm]{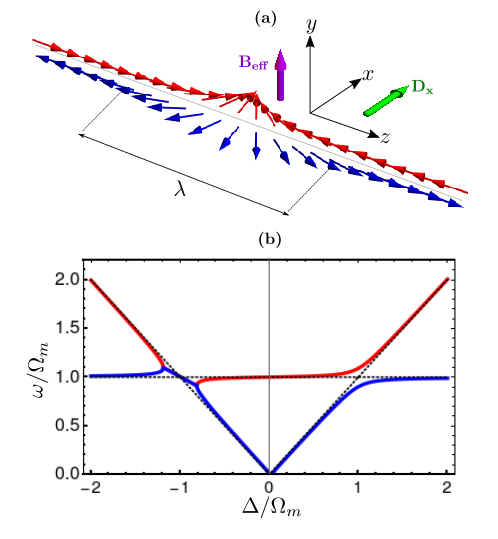}}
\caption{(a) Schematic illustrating the geometry where the effective magnetic field of the light along $\hat{y}$ interacts with the domain wall of width, $\lambda$ with axis along $\hat{z}$  in the presence of DMI along $\hat{x}$. (b) Hybridized frequency of the cavity modes and magnon modes of the antiferromagnetic domain wall as a function of the detuning parameter in the absence of DMI. The red and blue thick lines indicate coupling between the magnonic and photonic modes in the absence of DMI. The real part of the two modes attract in the region of negative detuning and repulsion is observed in the region of positive detuning. As the magnitude of DMI increases, coupling between modes reduces until no coupling is achieved. The black dotted lines show that there is no coupling at the saturated value of DMI.}
\label{fig:fig3}
\end{figure}

\section{Domain wall dynamics}
We now discuss effects of DMI on the dynamics of an antiferromagnetic domain wall when coupled to optical photons in an electromagnetic cavity.
We describe small amplitude oscillations by performing a Taylor series expansion of Eq.~(\ref{2}) about the equilibrium state $[\theta_0, \phi_0]$. Centering the fluctuations about the wall center at $Z_0 (t)$ \cite{vandermeulen,tveten2}, we describe these fluctuations with $ \phi (z,t) = \phi_0 (z- Z_0(t))$ and $\theta (z,t) = \theta_0 (z- Z_0(t)) + \delta \theta (z - Z_0(t))$. 
These excitations satisfy the P\"{o}schl-Teller equation. The solution, $ \delta \theta =P(t)~ \text{sech}[(z-Z_0(t))/\lambda]$, where $Z_0$ is the wall center and $P$ is the amplitude of the out-of-plane component.

In order to fully describe the dynamics of the antiferromagnetic domain wall, we introduce a Berry phase term, $\mathcal{L}_B = 2 \hbar S~ \vec{m}(\partial_t \vec{l} \times \vec{l})$ \cite{tveten,kim_2014}, which accounts for the inertia of the antiferromagnetic domain wall. A pinning potential provides the restoring force and is introduced as a point defect which contributes to the anisotropy along 
$\hat{y}$ at $z=0$. The pinned wall Hamiltonian with potential, $V_{\text{pin}} = -K_{\text{pin}}~ \delta(z)\sin^2 \theta (z) \sin^2 \phi (z) \approx (K_{\text{pin}}/\lambda^2) Z_0^2$ and the kinetic term upon integration, excluding terms that do not contribute to the equations of motion can then be written in terms of the wall position as:
\begin{equation}\label{3}
\begin{split}
\mathcal{H}=\frac{2 \rho^2 \dot{Z_0}^2}{a_0 \lambda} + \frac{K_{\text{pin}} Z_0^2}{2}
\end{split}
\end{equation}
where $\rho = 2 \hbar S$ and we have dropped terms proportional to the out-of-plane component, $P(t)$.

The coupling mechanism between the domain wall and cavity photon is assumed to be the Faraday effect \cite{faraday} in which the plane of polarization of light rotates when it passes through a magnetized medium. In this way, coupling appears as a reaction force to photon pressure exerted as compensation to changes in photon polarisation. Details can be found in Ref.~ \cite{proskurin_2019,kirilyuk,parvini,tzschaschel,silvia2018}. The magneto-optical interaction energy \cite{silvia2018} is expressed as
\begin{equation}\label{4}
\mathcal{H}_{\text{mo}} = - \frac{i}{4} f \epsilon_0 \int d^3r~ \vec{m}(\vec{r}) \cdot [\vec{E}^*(\vec{r},t) \times \vec{E}(\vec{r},t) ],
\end{equation}
where $\vec{m}$ is the total magnetization vector, $f$ is a Faraday rotation material-dependent parameter and $\vec{E}$ is the electric field vector of light. In this work, we express $\vec{m}$ in terms of the staggered magnetization, $\vec{l}$ since $\vec{m}$ is a slave variable which depends on the spatial variation of $\vec{l}$: $\vec{m}= \rho (\vec{l} \times \dot{\vec{l}}) + \vec{D} \times \vec{l}$.

The first of the two geometries we consider is that in which there is electromagnetic wave propagating along $\hat{x}$, and circularly polarized in $y$-$z$ plane. In this geometry, the resulting effective magnetic field of light in Eq.~(\ref{4}) is given by
 $\vec{E}^*(x) \times \vec{E}(x) = \hat{x} ~(i\hbar \omega)/(2V \epsilon_0) [\hat{a}^\dagger_R \hat{a}_R - \hat{a}^\dagger_L \hat{a}_L]$, where $V$ is the volume of the cavity and $\hat{a}_{R(L)}$ is related to the right (left) circularly polarized basis. The magneto-optical Hamiltonian obtained from Eq.~(\ref{4}) is $\mathcal{H}_{\text{mo}}^x$ = $f [\sin \phi \pi \dot{Z}_0 + 2 D_y Z_0]$, which shows that the interaction along $\hat{x}$ is proportional to the tilt angle of the domain wall and the $y$ component of the DMI (perpendicular to both the wall axis and the direction of propagation of electromagnetic waves). The magneto-optical interaction term is added to Eq.~(\ref{3}).

In addition, we introduce the photonic Hamiltonian, $\mathcal{H}_{\text{ph}}= \hbar \omega_c \hat{a}^\dagger \hat{a} $ and an external laser driving term, $\mathcal{H}_{\text{drive}}=\varepsilon \hat{a}^\dagger e^{-i \omega_d t} + \varepsilon^* \hat{a} e^{i \omega_d t}$, where $\varepsilon$ is the pump amplitude and $\omega_d$ is the driving frequency. From Eq.~(\ref{3}), we derive the canonical conjugate momentum $P_{Z_0}$  corresponding to $Z_0$, and the Hamiltonian is quantized by making the transformations: $\hat{Z}_0 = \sqrt{\hbar/2M \Omega_m} (\hat{b} + \hat{b}^\dagger)~ \text{and}~ \hat{P}_{Z_0} = -i \sqrt{\hbar M \Omega_m/2} (\hat{b}-\hat{b}^\dagger)$, where $\hat{b}^{(+)}$ are the annihilation (creation) operators of the magnonic mode. Employing the rotating frame approximation \cite{Aspelmeyer2014} to separate out the slow time dynamics and average over the fast, we move to a frame defined by the cavity field photon number rotating at a drive frequency $\omega_d$ by performing the unitary transformation $\tilde{H} = i (d\hat{U}/dt)+ \hat{U}~ \mathcal{H}_{\text{drive}}~ U^\dagger$, where $\hat{U}(t) = e ^{i\hbar \omega_d t \hat{a}^\dagger \hat{a}}$, then $\tilde{H} =  \hbar (\omega_d -\omega_c)\hat{a}^\dagger \hat{a} + \varepsilon \hat{a}^\dagger + \varepsilon^* \hat{a}$ and we arrive at:

\begin{multline}\label{5}
\mathcal{H} = \hbar \Omega_m \hat{b}^\dagger \hat{b}  + \hbar \Delta \hat{a}^\dagger \hat{a} + \varepsilon \hat{a}^\dagger + \varepsilon^* \hat{a} \\
- f \left[i \Omega_m \pi \sin \phi x_{\text{zpf}} (\hat{b}-\hat{b}^\dagger) - 2 D_y x_{\text{zpf}}(\hat{b}+\hat{b}^\dagger) \right] \hbar \omega_c \hat{a}^\dagger \hat{a},
\end{multline}
where $x_{\text{zpf}} = \sqrt{\hbar/(2 M \Omega_m)}$, the antiferromagnetic domain wall effective mass is $M=4 \hbar^2 /(a_0 K_z \lambda)$, the characteristic oscillation frequency, $\Omega_m = \sqrt{K_{\text{pin}}/(M \lambda^2)}$ and $\Delta= \omega_c -\omega_d$ is the wall oscillation detuning parameter.  For NiO having a domain wall width of approximately $150$ nm \cite{tzschaschel}, $a_0 = 0.418$ nm \cite{peck} and $K_z= 4~ \text{K}$ \cite{kondoh}, we assume $K_{\text{pin}}= 1 \text{K}$ such that it is of the same order of magnitude as the anisotropy constant. This gives $M \approx 10^{-29} kg$ and $\Omega_m \approx~ 10~ \text{GHz}$.

The equations of motion for $\hat{a}$, $\hat{a}^\dagger$, $\hat{b}$ and $\hat{b}^\dagger$ in the Appendix are obtained from Eq.~(\ref{5}) and linearized by splitting the operators into an average and a fluctuating term. e.g. $\hat{a}= \langle \hat{a} \rangle + \delta \hat{a}$, where $\langle \hat{a} \rangle$ is related to the average number of cavity photons, $\bar{n}_c =|\langle \hat{a}\rangle|^2$. Solving the equations of motion for the hybridized frequency $\omega$, we obtain two pairs of eigenmodes:
\begin{equation}\label{6} 
\omega_{\pm} = \sqrt{\frac{\Delta^2 + \Omega_m^2}{2} \pm \sqrt{\frac{(\Delta^2 -\Omega_m^2)^2}{4}-16 g^2 D_y^2 \Delta \Omega_m }},
\end{equation}
where $g$ is the measure of the coupling of magnons to photonic modes. The coupling strength $g$ is often expressed in terms of the extensive macrospin, S and the Faraday rotation, $\theta_F$~ \cite{proskurin_2019,viola2016,silvia2018,parvini}~: $g= (c~\theta_F/4 \sqrt{\varepsilon S})$, where $c$ is the speed of light and $\varepsilon$ is the dielectric constant of the material. We estimate $g\approx 0.01$ MHz for a $100~\mu$m-thick NiO slice with $\theta_F=$ 30 rad m$^{-1}$~ \cite{tsatoh}, $\varepsilon =11.9$ \cite{rao} and $S \approx10^{10}$. This coupling strength is one order of magnitude less than what is obtained for YIG \cite{viola2016}.
Equation (\ref{6}) shows that if there were no coupling (i.e $g=0$), there would be crossing of the modes. The same result is realized in the absence of DMI since $D_y = 0$ and $\phi(D_y) = 0$ give no room for the coupling in Eq. (\ref{6}) . The dotted lines in Fig (\ref{fig:fig1}~b) show crossing of modes in both regions ($\Delta >0$ and $\Delta<0$) when DMI is absent, and no coupling is achieved. The thick lines show that in the presence of DMI, degeneracy is broken and avoided crossing of modes is realized for positive detuning ($\Delta >0$), indicating coherent coupling of the modes. In the region of negative detuning ($\Delta < 0$), there is an exceptional point where the eigenfrequencies become complex and there is coalescence of their real parts resulting in attraction of the modes. 
\begin{figure*}
\centering
\includegraphics[width=1.0\textwidth]{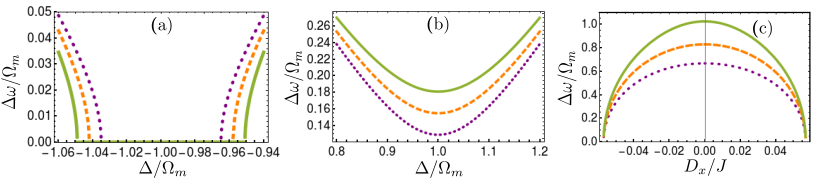}
\caption{The effect of DMI of strength $0.06 J$ on the frequency gaps as (a) a function of the negative detuning around the point of attraction for different coupling strengths: $g=0.10$, $g=0.12$, and $g=0.14$ represented by the purple dotted, orange dashed and green solid lines, respectively. Around the exceptional point, attraction dominates a wider range for larger coupling strengths and decreases for lower coupling strengths.  (b) Frequency gap in the presence of DMI of strength $0.06 J$ as a function of the positive detuning. For the same value of detuning, larger frequency gaps are observed for larger coupling. In all cases, the gap is minimum at resonance and increases as we move away from resonance. (c) Frequency gap as a function of negative and postive DMI present along the $\hat{y}$. The values of the coupling strength $g$ become more important as $D_x$ approaches zero.  The frequency gap reduces for increasing magnitude of the DMI and vanishes as the saturated value of DMI is reached.}
\label{fig:fig4}
\end{figure*}

In order to understand the consequence of the presence of DMI on resonance, we study frequency gaps, $\Delta \omega = \omega_+ - \omega_-$. In Fig.~(\ref{fig:fig2}), $\Delta \omega$ is shown as a function of $\Delta$ for different values of $g$. In the region of negative detuning (Fig \ref{fig:fig2}~a), around the exceptional point, there is no frequency gap and the attraction regime spans  a wider range of detuning for increasing coupling strengths. Away from the exceptional point we observe a difference in frequencies of the two modes. On the other hand, in the region of repulsion (Fig.~\ref{fig:fig2}~b), we observe that at near crossing, a frequency gap exists. This gap can be seen to increase with increasing coupling strength: In Fig. (\ref{fig:fig2}~c), we describe how frequency gaps for positive and negative $\Delta$ depend on DMI at resonance. The gaps clearly increase with increasing magnitude of DMI and there is no gap in the absence of DMI, i.e., the modes then cross. We conclude that the coupling of the photonic mode to the magnonic mode is dependent on the DMI strength, and is proportional to the coupling strength in this geometry and there will be no coupling in the absence of DMI.

The second geometry considered in this work is the case where the domain wall interacts with waves propagating along $\hat{y}$, and circularly polarized in the $x$-$z$ plane (Fig \ref{fig:fig3}~a). Whether DMI is present or not, coupling of cavity modes to the domain wall excitations is achieved. We observe that this coupling is stronger than in the first geometry considered due to the fact that the electromagnetic wave is applied in a direction perpendicular to the easy plane of the domain wall, thus maximizing coupling to variation of magnetization. Following a similar approach described for the first geometry considered, we obtain two pairs of eigenmodes: 
\begin{widetext}
\begin{equation}\label{7} 
\omega_{\pm} = \sqrt{\frac{\Delta^2 + \Omega_m^2}{2} \pm \sqrt{\frac{(\Delta^2 -\Omega_m^2)^2}{4}-16 g^2 D_x^2 \Delta \Omega_m-2 g^2 \pi^2 \Delta \Omega^3 (1 + \cos 2 \phi) }}    
\end{equation}
\end{widetext}
Frequencies from Eq.~(\ref{7}) are shown in Fig. (\ref{fig:fig3}~b), where the hybridized frequency is plotted as a function of the detuning parameter $\Delta$. First, we discuss what happens when DMI is absent ($D_x = 0$). The thick red and blue lines in Fig. (\ref{fig:fig3}~b) indicate that there is coupling even in the absence of DMI.

However, when DMI is present along $\hat{x}$, the spin orientations are distorted out of the plane of the wall and we have a non-zero tilt angle which depends on the DMI strength. This modifies the domain wall width as $\lambda_{D_x}=\sqrt{A/(K_z - K_x (8 J K_x^2 - D_x^2 K_z \pi^2)/(8 J K_x^2 - D_x^2 K_x \pi^2))}$. The consequence of this is weaker magneto-optical coupling in the  interaction for both positive and negative values of DMI, and symmetry about $D=0$ up to the saturated value, $D_{\text{sat}} = 4/\pi \sqrt{J K_x^2/2 K_z}$ (when crossing of the modes emerges in both regions). We see how DMI reduces the coupling between modes until crossing of the mode is achieved just as $D= D_{\text{sat}}$ as indicated by the black dotted lines in Fig.(\ref{fig:fig3}~b).

The effects of DMI on resonance frequencies are observed in the region surrounding the exceptional points. The dependence of the frequency gaps on the detuning in the presence of DMI along $x$ is similar to what we observed along $y$ except that the attraction and repulsion are stronger for the same value of DMI if we compare Figs. (\ref{fig:fig2}~a) and (\ref{fig:fig2}~b) to Fig. (\ref{fig:fig4}~a) and (\ref{fig:fig4}~b). The most striking feature that distinguishes both geometries is how the frequencies change with respect to the DMI. Contrary to the previous geometry, increase in the DMI applied along $x$ leads to decrease in the frequencies until the gap vanishes at the saturated value of DMI (Fig. \ref{fig:fig4}~c).

It is useful to define the frequency gap $\Delta \omega(D_x)$ in order to see clearly how the frequency gap depends on DMI at resonance. We calculate this from Eq.~(\ref{7}): $ \Delta \omega(D_x)= \omega_+ - \omega_-|_{\scriptscriptstyle{\Omega_m=\Delta=1}} \approx 2g (8 D_x^2 + \pi^2 (1 + \cos 2 \phi(D_x)))$. For example, if we assume a coupling strength of 0.1 at a tilt angle of $\pi/2$, then $\Delta \omega \approx 2 D_x^2$. This shows that the gap depends on the DMI constant and may be used to determine the DMI strength for given values of the coupling strength.

\section{Conclusion}
We demonstrated that collective excitation of a Ne\'{e}l antiferromagnetic DW can be realized through the magneto-optical coupling to cavity photons.  The resulting Hamiltonian is of an optomagnonic type, which allows realization of optomechanical instabilities such as level attraction in an driven system. The coupling strength of cavity photons to magnons in antiferromagnetic NiO were estimated and found to be an order of magnitude lower than that of YIG. We find that the presence of DMI enables a coupling between DW and cavity photons in a geometry where there is no coupling otherwise. This opens a possibility for estimating DMI in antiferromagnetic materials by measuring the interaction of antiferromagnetic resonances to optical modes in a microwave cavity. This approach is not limited to a single DW dynamics but applicable to other one dimensional textures like chiral soliton lattice in ferromagnets and antiferromagnets and can, in principle, be extended to antiferromagnetic spin textures in two spatial dimensions such as skyrmions and skyrmion lattices, which remains as a future problem.\\

\section{Acknowledgements}
\begin{acknowledgements}
This work was supported by The Natural Sciences and Engineering Research Council of Canada (NSERC) Discovery, John R. Leaders Fund - Canada Foundation for Innovation (CFI-JELF), Research Manitoba and the University of Manitoba, Canada.
\end{acknowledgements}

\bibliography{main}

\newpage
\onecolumngrid

\appendix*
\section{APPENDIX}
\subsection{Dzyalonshinskii-Moriya interaction}
\label{APP:DMI} 
The DMI between two atomic spins $S_i$ and $S_{i+1}$ can be expressed as vector product formed by the magnetic moments $S_i$ of two magnetic ions,
$$\mathcal{H}_{\text{DMI}} = \sum_i (-1)^i \vec{D} \cdot (\vec{S}_i \times \vec{S}_{i+1} )$$
 $$\mathcal{H}_{\text{DMI}} = \sum_n \vec{D} \cdot (\vec{S}_n^A \times \vec{S}_n^B) + \vec{D}(\vec{S}^A_{n+1} \times \vec{S}_n^B) $$
where $\vec{S}^A = \vec{m} + \vec{l}$ and $\vec{S}^B = \vec{m} - \vec{l}$. Substituting these into the previous expression and doing some lines of vector algebra, we arrive at the expression:
$$\mathcal{H}_{\text{DMI}} = \sum \left\lbrace 4 \vec{D} S^2 \cdot (\vec{m}_n \times \vec{l}_n)- \vec{D} S^2 [\vec{m}_{n+1} \times \vec{m}_n - \vec{l}_{n+1} \times \vec{l}_n - (\vec{m}_{n+1} - \vec{m}_n) \times \vec{l}_n + (\vec{l}_{n+1} - \vec{l}_n) \times \vec{m}_n ]\right\rbrace $$
In the continuum limit,
\begin{equation}\label{A1}
\mathcal{H}_{\text{DMI}} = \int \frac{dz}{a_0}  \vec{D} \cdot \left(\frac{\partial \vec{l}}{\partial z} \times \vec{l} \right)
\end{equation}

\subsection{DMI-dependent equations of motion}
\label{APP:Eqn_motion}  
The equations of motion in the presence of DMI are given. First, we consider the case where electromagnetic waves propagating along $\hat{x}$ couple to the wall position with DMI present in the $\hat{y}$ direction. The equations of motion in frequency space
  \begin{equation}\label{B1}
   \setlength{\arraycolsep}{0.5pt}
   \renewcommand{\arraystretch}{1.0} \begin{pmatrix} 
   -i (\omega- \Omega_m) &0 &-g(\pi \Omega_m \sin \phi-2i D_y) & -g(\pi \Omega_m \sin \phi-2i D_y)\\
   0& -i(\omega + \Omega_m)&-g(\pi \Omega_m \sin \phi+2i D_y) & -g(\pi \Omega_m \sin \phi+2i D_y)\\
   g(\pi \Omega_m \sin \phi+2i D_y) &-g(\pi \Omega_m \sin \phi-2i D_y)& -i(\omega - \Delta)&0  \\
   -g(\pi \Omega_m \sin \phi+2i D_y)& g(\pi \Omega_m \sin \phi-2i D_y) &0 & -i(\omega+\Delta)
   \end{pmatrix} 
   \setlength{\arraycolsep}{0.5pt}
   \renewcommand{\arraystretch}{1.0}  \begin{pmatrix} 
   \delta \hat{b}  \\
   \delta \hat{b}^\dagger\\
   \delta \hat{a}  \\
   \delta \hat{a}^\dagger
   \end{pmatrix}=
   \begin{pmatrix} 
   0 \\
   0\\
   -i \varepsilon\\
   i \varepsilon*
   \end{pmatrix}
\end{equation}
In this configuration, the propagating wave is in the hard axis of the wall and results in no magneto-optical coupling. We found that for DMI to play a role, it must be perpendicular to both the wall direction ($\hat{z}$) and the direction of propagating wave ($\hat{x}$). Therefore, we consider DMI present along $\hat{y}$. The presence of DMI breaks the degeneracy which occurs at resonance and allows for magneto-optical coupling which depends on the y-component of the DMI vector.

The second geometry considered is the one in which  electromagnetic wave propagating along $\hat{y}$ couples to the wall position with DMI present in the $\hat{x}$ direction. The electromagnetic wave is circularly polarized in the easy plane of the domain wall. As a result, there is a strong coupling. The plot of hybridized frequency against the detuning parameter is shown in Fig (\ref{fig:fig3}b). We draw the conclusion that in this geometry whether DMI is present or not, there is coupling of photons to magnons, however the presence of DMI leads to modulation of the frequencies. The equation of motion is given in matrix form:

\begin{equation}\label{B2}
   \setlength{\arraycolsep}{0.5pt}
   \renewcommand{\arraystretch}{1.0} \begin{pmatrix} 
   -i (\omega- \Omega_m) &0 &g(\pi \Omega_m \cos \phi-2i D_x) & g(\pi \Omega_m \cos \phi-2i D_x)\\
   0& -i(\omega + \Omega_m)&g(\pi \Omega_m \cos \phi+2i D_x) & g(\pi \Omega_m \cos \phi+2i D_x)\\
   -g(\pi \Omega_m \cos \phi+2i D_x) &g(\pi \Omega_m \cos \phi-2i D_x)& -i(\omega - \Delta)&0  \\
   g(\pi \Omega_m \cos \phi+2i D_x)& -g(\pi \Omega_m \cos \phi-2i D_x) &0 & -i(\omega+\Delta)
   \end{pmatrix} 
   \setlength{\arraycolsep}{0.5pt}
   \renewcommand{\arraystretch}{1.0  }  \begin{pmatrix} 
   \delta \hat{b} \\
   \delta \hat{b}^\dagger\\
   \delta \hat{a}\\
   \delta \hat{a}^\dagger
   \end{pmatrix}=
   \begin{pmatrix} 
   0 \\
   0\\
   -i \varepsilon\\
   i \varepsilon*
   \end{pmatrix}
  \end{equation}

\end{document}